\documentstyle[twoside,fleqn,espcrc2]{article}

\newcommand{\AmS}{{\protect\the\textfont2
  A\kern-.1667em\lower.5ex\hbox{M}\kern-.125emS}}
\newcommand{\real}{{\sf I}\kern-.12em{\sf R}}

\title{Three-Loop Results in QCD with Wilson Fermions}

\author{B. All\'es\address{Dipartimento di Fisica and I.N.F.N., Milano, Italy}
        ,
        C. Christou\address{Department of Natural Sciences, University
        of Cyprus}
        ,
        A. Feo$^{\rm b,}$\address{Dipartimento di Fisica, Universit\`a di Pisa, Italy}
        ,
        H. Panagopoulos$^{\rm b}$\thanks{Presented the paper.} and
        E. Vicari$^{\rm c,}$\address{I.N.F.N., Sezione di Pisa, Italy}}

\begin{document}

\begin{abstract}
We calculate the third coefficient of the lattice beta 
function in QCD with Wilson fermions, extending the pure
gauge results of L\"uscher and Weisz; we show how this 
coefficient modifies the scaling function on the lattice.
   We also calculate the three-loop average plaquette in
the presence of Wilson fermions. This allows us to compute
the lattice scaling function both in the standard and 
energy schemes.
\end{abstract}

\maketitle

\section{INTRODUCTION}

In lattice formulations of field theories
the relevant region 
for continuum physics is the one where scaling is verified.
Scaling in the continuum limit requires that
all RG-invariant dimensionless ratios of physical quantities 
approach their continuum value with
non-universal corrections which get depressed
by integer powers of the inverse correlation length.
Tests of asymptotic scaling involve the relationship between the bare coupling and the
cutoff as an essential ingredient. This relationship is expressed in
the bare lattice $\beta$ function which reads, in standard notation:
\begin{equation}
\beta^L(g_0) {\equiv} {-} a \frac{\hbox{d}g_0}{\hbox{d}a} \mid_{g_r,\mu}=
{-} b_0^L g_0^3 {-} b_1^L g_0^5 {-} b_2^L g_0^7 - \ldots
\end{equation}
The first two coefficients in Eq. (1) are scheme-independent and well
known. 
Our purpose is to compute the third coefficient,
$b_2^L$, in $SU(N)$ Yang-Mills theory with $N_f$ species of Wilson fermions.

M. L\"uscher and P. Weisz  \cite{lw234} have calculated this
coefficient in the pure gluonic case,
using a coordinate space method
for evaluating the lattice integrals. We have verified their result \cite{afp},
using a different technique where firstly the integrands are Taylor expanded
in powers of the external momenta and then computed term by term with
the introduction of an IR regulator. In the present work, we are
extending our calculation to include contributions from Wilson fermions.

We also present here the calculation of the 3-loop coefficient $w_3$ of the average
plaquette, in QCD with Wilson fermions. In the absence of fermions,
this quantity was
calculated by some of the present authors \cite{plb}.

Our results for $b_2^L$ and $w_3$ allow us to study the corrections to
asymptotic scaling in Monte Carlo simulations of 
dimensionful quantities, for both the standard and energy schemes.

The involved algebra of lattice perturbation theory was
carried out by making use of an extensive computer code developed by
us in recent years \cite{npb}; for the purposes of the present calculation, the
code was extended to include fermions and form factors.

\section{CALCULATION OF $\beta_2^L$}

We performed this calculation using the background field method. In order
facilitate comparison we have adopted the notation of
Ref.\cite{lw234}.

We set out to compute $d_2(\mu a)$ in:
$$\alpha_{\overline{\rm MS}}(\mu) = \alpha_0 + d_1(\mu a) \alpha_0^2 + 
   d_2(\mu a) \alpha_0^3 + \ldots,$$
\begin{equation}
\quad\alpha_0 = g_0^2/4\pi, \quad \alpha_{\overline{\rm MS}}(\mu) = g^2(\mu)/4\pi
\end{equation}
As we shall see, $b_2^L$ will follow immediately from this
quantity. $d_2(\mu a)$ can be expressed as:
$$d_2(\mu a){=} (4\pi)^2 \{ (\nu_R^{(1)}(p) - \nu^{(1)}(p))^2 -
\nu_R^{(2)}(p) + \nu^{(2)}(p) $$
\begin {equation}
\qquad - \lambda (\partial \nu_R^{(1)}(p)/\partial\lambda) 
(\omega^{(1)}(p) - \omega_R^{(1)}(p))\}_{\lambda = \lambda_0}
\end{equation}
in terms of the gauge parameter $\lambda$, and of $\omega(p)$ and
$\nu(p)$ (the gauge- and background- two point lattice functions, respectively):
$$\sum_\mu \Gamma^{(2,0,0)}(p,-p)_{\mu\mu}^{ab} {=} - \delta^{ab} 3 \hat
p^2 [1-\nu(p)]/g_0^2$$
$$\sum_\mu \Gamma^{(0,2,0)}(p,-p)_{\mu\mu}^{ab} {=} - \delta^{ab} \hat
p^2 [3(1-\omega(p)) + \lambda_0]$$
\begin{eqnarray}
&&\nu(p) =\sum_{l=1}^\infty g_0^{2l}\nu^{(l)}(p)\nonumber\\
&&\omega(p) =\sum_{l=1}^\infty g_0^{2l}\omega^{(l)}(p)
\end{eqnarray}
The subscript $R$ refers to the analogous $\overline{\rm MS}$ renormalized
quantities.

The 1-loop quantitites appearing above, $\nu^{(1)}, \nu_R^{(1)},
\omega^{(1)}, \omega_R^{(1)}$, are known. For the $\overline{\rm MS}$
2-loop function $\nu_R^{(2)}(p)$ there are four new diagrams with
respect to the pure gluonic case. We find:
$$
\nu_R^{(2)}(\lambda{=}1,\,p) (16\pi^2)^2 = N^2 [8\rho
+577/18-6\zeta(3)]+
$$
$$\,\,N_f[(-3\rho-401/36)N+(\rho+55/12-4\zeta(3))/N]$$
\begin{equation}
{\rm with:}\quad \rho = \ln(\mu^2/p^2).
\end{equation}

 The only remaining quantity to calculate
(and by far the most complicated) is $\nu^{(2)}(p)$.
The inclusion of fermions brings in 20 additional diagrams (of these,
two correspond to fermion mass renormalization).

The evaluation of these diagrams requires tremendous analytical
effort. We have developed an extensive computer code for analytic
computations in lattice perturbation theory \cite{npb}. In a nutshell,
this computer code performs: Contraction among the appropriate
vertices; simplification of color/Dirac matrices; use of trigonometry
and momentum symmetries for reduction to a more compact, canonical
form; treatment of (sub)divergences and extraction of logarithms by
opportune additions/subtractions (in the case at hand, this step gives
rise to hundreds of {\it types} of expressions, each containing
typically hundreds of terms); automatic generation of highly optimized
Fortran code for the loop integration of each type of expression.

The integrals are then performed numerically on finite lattices. Our
programs perform extrapolations of each expression to a broad spectrum
of functional forms of the type: $\sum_{i,j} e_{ij} (\ln L)^j/L^i$,
analyze the quality of each extrapolation using a variety of criteria
and assign statistical weights to them, and finally produce a quite
reliable estimate of the systematic error.

Several consistency checks can be performed on the separate
contributions of each diagram: ${\cal O}(p^0)$ parts obey several relations, and
must sum up to zero by gauge invariance; Lorentz non-invariant terms
$\sum_\mu p_\mu^4/p^2$ must cancel; the single and double logarithms
are related to known results. We have performed all these checks both
analytically and numerically; this was also a useful verification of
the correctness of our error estimates.

Our final result for $\nu^{(2)}(p)$ is (we take $r=1$ for the Wilson parameter):
\begin{equation}
\nu^{(2)}(p) = \nu^{(2)}(p)|_{N_f=0} 
\end{equation}
\begin{eqnarray}
&&+ N_f [(-1/N + 3N)
\ln(p^2)/(16\pi^2)^2\nonumber\\
&& - 0.000706(3) /N + 0.000544(5) *N ]\nonumber
\end{eqnarray}
Also, we have: 
$$\nu^{(1)}(p) - \nu^{(1)}(N_f{=}0,p) = \omega^{(1)}(p) -
\omega^{(1)}(N_f{=}0,p)$$
\begin{equation}
\quad = N_f * [ \ln(p^2) /24\pi^2 - 0.01373219(1)]
\end{equation}

A table of partial results will appear in a forthcoming publication
\cite{cfpv}, where we also expect to provide better accuracy.

 From the relations given above, we can now easily construct $d_1(\mu
a)$ and $d_2(\mu a)$. Writing: $d_1(\mu a) = d_{10} + d_{11} \ln(\mu
a), \quad d_2(\mu a) = (d_1(\mu a))^2 + d_{20} + d_{21} \ln(\mu a)$,
the coefficient $b_2^L$ follows immediately:
\begin{equation}
b_2^L = b_2^{\overline{\rm MS}} +  (d_{20} d_{11} -
d_{21} d_{10})/(128\pi^3)
\end{equation}
where
$$b_2^{\overline{\rm MS}} = {-1\over 108 (4\pi)^6} [ - 5714
N^3 + N_f^2 (-224 N + 66/N) $$
\begin{equation}\qquad\qquad + N_f (3418 N^2 - 561 - 27/N^2) ]
\end{equation}

In particular:
\begin{equation}
b_2^L(N{=}3,N_f{=}3,r{=}1) = -0.002284(3) 
\end{equation}
\begin{equation}
({\rm cf.}\qquad b_2^L(N{=}3,N_f{=}0) = -0.00159983)
\end{equation}

The deviation, $q = (b_1^2 - b_2^L b_0)/(2b_0^3)$, in the asymptotic
scaling formula:
$$a\Lambda_L = \exp(-1/ 2b_0g_0^2)
(b_0g_0^2)^{-b_1/ 2b_0}
\left[ 1 {+} q g_0^2 {+} \ldots\right] $$
is now found to be: 
\begin{equation}
q(N{=}3,N_f{=}3,r{=}1) = 0.3694(4)
\end{equation}

\section{THE 3-LOOP MEAN PLAQUETTE}

We have computed the free energy to 3 loops 
\begin{equation}
-\ln Z/V = {\rm const.} + F_2 g_0^2 + F_3 g_0^4 + \ldots
\end{equation}
in QCD with Wilson fermions. A total of 24 diagrams involving fermions
contribute to $F_3$.

The equality:
\begin{equation}
\langle 1 - \Box/N\rangle = -(1/6V)(\partial \ln Z /
\partial\beta)
\end{equation}
 (valid by virtue of $\langle S_{\rm fermion}\rangle = 0$ beyond
tree level) relates $F_2$ and $F_3$ directly to the quantities $w_2$
and $w_3$ in the expansion of the average plaquette:
\begin{equation}
\langle 1 - \Box/N\rangle = w_1 g_0^2 + w_2 g_0^4 + w_3 g_0^6 +
\ldots \end{equation}
We report below the final values for 
$w_2$ and $w_3$ at $N{=}3, N_f{=}3, r{=}1$.
A breakdown of the results, along with the explicit
$N$- and $N_f$-dependence and improved error estimates will be
presented in a forthcoming publication \cite{afp2}.

 From the plaquette coefficients, we may now construct the coupling
constant in the ``energy'' scheme:
\begin{equation}
g_E^2 = g_0^2 + (w_2/w_1) g_0^4 + (w_3/w_1) g_0^6 + \ldots
\end{equation}
In terms of this improved definition of the coupling constant, the
$\beta$ function reads:
\begin{equation}
b_2^E = b_2^L + (b_0 w_3 - b_1 w_2 - b_0 w_2^2/w_1)/w_1 
\end{equation}
Now, the scaling violation parameter $q = (b_1^2 - b_2^E
b_0)/(2b_0^3)$ becomes (at $N=3, N_f=3, r=1$):
\begin{eqnarray}
&&q(k=0.1675)= 0.2103(4)\nonumber\\
&&q(k=0.1560)= 0.1764(4)
\end{eqnarray}
This indeed represents an improvement over the standard case.
We are presently working on improving the accuracy of our results, for
the longer write-ups \cite{cfpv,afp2}.

\begin{table}[h]     
\begin{center}
\begin{tabular}{|c|c|c|}
\hline
\hline
$k$ & $10^3 \times w_2$ & $10^3 \times w_3$ \\
\hline
\hline
0.156 &
                18.2522(2)
                               &
                  9.15(1) \\
0.1575 &
                 17.7713(2)        
                               &
                        8.91(1) \\
0.16 &
                    16.9767(1)   
                               &
                       8.51(1) \\
0.164 &
                15.7243(2)
                            &
                         7.89(1) \\
0.1675 &
                 14.6480(6)
                               &
                     7.35(1) \\
\hline
\hline
\end{tabular}
\caption{Values of $w_2 \times 10^{3}$ and $w_3 \times
10^{3}$. $N{=}3$, $N_f{=}3$, $r{=}1$, $k$: hopping parameter.}
\label{...}
\end{center}
\end{table}

\end{document}